\documentclass[authoryear,12pt,5p,times]{elsarticle}

\usepackage{aas_macros}
\usepackage{epsfig}
\usepackage{verbatim, amsmath, amsfonts, amssymb,amssymb}
\usepackage[utf8x]{inputenc}
\usepackage[table]{xcolor}
\usepackage{booktabs, longtable}
\usepackage{multirow}
\usepackage{float}
\usepackage{lscape}
\usepackage{graphicx, subfigure}
\usepackage{rotating}
\usepackage[colorinlistoftodos]{todonotes}
\usepackage{animate}
\usepackage{natbib}
\usepackage{enumerate}
\usepackage{url}
\usepackage{bm}
\usepackage{tikz}
\usepackage{color}
\usepackage[lined]{algorithm2e}
\usepackage{listings}
\usepackage{animate}
\usepackage{adjustbox}
\usepackage{textcomp}

\begin{document}

\pagestyle{myheadings}
\markboth{Likelihood-free inference via PMC-ABC}{Likelihood-free inference via PMC-ABC}

\title{cosmoabc: Likelihood-free inference via Population Monte Carlo Approximate Bayesian Computation}
\author[Ishida et al.]
{E. E. O. Ishida$^{1}$, S. D. P. Vitenti$^{2}$,  M. Penna-Lima$^{3,4}$, J. Cisewski$^{5}$, R. S. de Souza$^{6}$, A. M. M. Trindade$^{7,8}$ \\E. Cameron$^{9}$ and V. C. Busti$^{10,11}$, for the COIN collaboration\\
\small{$^{1}$Max-Planck-Institut f{\"u}r Astrophysik, Karl-Schwarzschild-Stra\ss e 1, 85748, Garching, Germany\\
$^{2}$ ${\mathcal G}\mathbb{R}\varepsilon\mathbb{C}{\mathcal O}$ -- Institut d'Astrophysique de Paris, UMR7095 CNRS, Universit\'e Pierre \& Marie Curie, 98  \\ bis boulevard Arago, 75014 Paris, France\\
$^{3}${APC, AstroParticule et Cosmologie, Universit\'e Paris Diderot, UMR7164 CNRS/IN2P3, \\10 rue Alice Domon et L\'eonie Duquet, 75013 Paris, France}\\
$^{4}${Instituto Nacional de Pesquisas Espaciais, Divisão de Astrofísica,
Av. dos Astronautas 1758, \\12227-010, São José dos Campos – SP, Brazil}\\
$^{5}$Department of Statistics, Yale University, New Haven, CT,  06511, USA\\
$^{6}$MTA E\"otv\"os University, EIRSA ``Lendulet'' Astrophysics Research Group, Budapest 1117, Hungary\\
$^{7}$Instituto de Astrofisica e Ciências do Espaço, Universidade do Porto, CAUP, Rua das Estrelas, \\ PT4150-762 Porto, Portugal\\
$^{8}$Departamento de Física e Astronomia, Faculdade de Ciências, Universidade do Porto, \\
Rua do Campo Alegre 687, PT4169-007 Porto, Portugal\\
$^{9}$ Department of Zoology, University of Oxford, Tinbergen Building, \\ South Parks Road, Oxford, OX1 3PS, United Kingdom\\
$^{10}$ Astronomy, Cosmology and Gravity Centre (ACGC), Department of Mathematics and Applied Mathematics, \\
University of Cape Town, Rondebosch 7701, Cape Town, South Africa \\
$^{11}$ Departamento de Física Matemática, Instituto de Física, Universidade de São Paulo, CP 66318, \\
CEP 05508-090, São Paulo - SP, Brazil}}

\label{firstpage}


\topmargin -1.3cm


\begin{abstract}
Approximate Bayesian Computation (ABC) enables parameter inference for complex physical systems in cases where the true
likelihood function is unknown, unavailable, or computationally too expensive.
It relies on the forward simulation of mock data and comparison between observed and synthetic catalogues.  
Here we present \textsc{cosmoabc}, a Python ABC sampler featuring a {\it Population Monte Carlo} variation of the original ABC algorithm, which uses an adaptive importance sampling scheme. The code is very flexible and can be easily coupled to an external  simulator, while allowing to incorporate arbitrary distance and prior functions. 
As an example of practical application, we coupled \textsc{cosmoabc} with the \textsc{numcosmo} library 
and demonstrate how it can be used to estimate posterior probability distributions over cosmological parameters based on measurements of galaxy clusters number counts without  computing the likelihood function.
\textsc{cosmoabc} is published under the GPLv3 license on PyPI and GitHub and documentation is available at \url{http://goo.gl/SmB8EX}.
\end{abstract}
\maketitle


\section{Introduction}
\label{sec:intro}

The precision era of cosmology marks the transition from a data-deprived field to a data-driven science on which statistical methods play a central role. The ever-increasing  data deluge  must be tackled  with new  and innovative statistical methods in order to improve our understanding of the key ingredients driving  our Universe \citep[e.g.,][]{Borne2009,Ball2010,deSouza2014a,deSouza2015}.
Given the continuous  inflow of new data, one does not start an analysis from scratch  for  every new telescope, but is
guided by previous  knowledge accumulated through  experience. A new experiment provides extra information  which needs to be incorporated into the larger picture, representing a small update on the previous body of knowledge. Such a learning process is
a canonical scenario to be embedded in a Bayesian framework, which allow us to
update our degree of belief on a set of model parameters\footnote{For the purposes of this work, we will only be interested in the \textit{parameter values} of a given model. However, it is important to stress that in a completely Bayesian approach all the elements and hypotheses  forming the model can be considered part of the \textit{prior}. In this sense, with the arrival of new essential information, the Bayesian approach allows for completely redefinition of the model itself \citep{kruschke2011}.} whenever  new and independent data are acquired.

A standard Bayesian analysis specifies prior distributions on unknown parameters, defines 
which parameter values better describe the relationship between the model, the prior  and the new data, and then finds the posterior
distribution -- either analytically or via sampling techniques, as e.g. with  Markov Chain Monte Carlo  \citep[MCMC; ][]{Metropolis1953}. This 
analysis requires a proper construction of the  likelihood function, which is not  always well known or easy to handle.   A common solution
would be to construct a model for the  likelihood (e.g. a Gaussian) followed by MCMC, with the expectation that
this hypothesis  is not too far from the true.  
Nonetheless, the challenge of performing parameter inference 
from an unknown or intractable   likelihood function   is becoming  familiar to
the modern astronomer.  
Recent  efforts to overcome observational
selection biases in the study of massive \citep{sana2012} and not-so
massive \citep{janson2014} stars, to account for windowing effects,
errors and/or gaps in time-series of X-ray emission from active galactic
nuclei \citep{uttley2002,shimizu2013} and UV emission from stellar coronae
\citep{kashyap2002} have been reported.

The development of \textit{ad hoc} approaches to this
problem within astronomy have proceeded independently from their long history within the field of
population genetics. The latter have ultimately been formalized into a rigorous
statistical technique known as Approximate Bayesian Computation (ABC).
The intuition of ABC dates back to a thought-experiment in \cite{Rubin1984}, where the basic ABC rejection sampler is used to illustrate Bayes Theorem.  \citet{tavare1997} employs an acceptance-rejection method in the context of population genetics, while \citet{pritchard1999} presents the first implementation of a basic ABC  algorithm.  Only recently has
the ABC approach been introduced and applied to astronomical problems
\citep{cameron2012, schafer2012, weyant2013,robin2014}. This work is part of a larger endeavour  natural consequence of such initial efforts. Following the philosophy behind the \textit{Cosmostatistics Initiative} (COIN)\footnote{\url{http://goo.gl/rQZSAB}}, we present a tool which enables  astronomers to easily introduce ABC techniques into their daily research.

The cornerstone of the ABC approach is our capability of performing quick and reliable computer simulations which mimic the observed data in the best possible way (this is called \textit{forward simulation inference}). In 
this context, our task relies on performing a large number of simulations and quantifying the ``distance'' between the simulated and observed catalogues. The better a parametrization reproduces the observed data in a simulated context, the closer it is to the ``true'' model. From this simple reasoning, many alternatives were developed  to optimize the parameter space sampling and the definition of distance. One of such examples is the work of \citet{marjoram2003}, who proposes a merger between the standard MCMC algorithm and the ABC rejection sampling. In astronomy, the method was used by \citet{robin2014} to constrain the Milky Way thick disk formation. Going one step further, \citet{BeaumontEtAl2009} propose to evolve an initial set of parameter values (or \textit{particle system}) through incremental approximations to the true posterior distribution. The \textit{Population Monte Carlo ABC} (PMC-ABC), method was used to make inferences on rate of  morphological transformation  of galaxies at high redshift \citep{cameron2012} and proved to be efficient in tracking the Hubble parameter evolution from type Ia supernova measurements,  despite the contamination from type II supernova  \citep{weyant2013}. More recently, \citet{lin2015} used ABC to predict weak lensing peak counts and  \citet{killedar2015} applied a weighted variant of the algorithm to cluster strong lensing cosmology.

This work introduces {\sc cosmoabc}\footnote{https://pypi.python.org/pypi/CosmoABC}, the first   publicly available\footnote{Shortly after \texttt{cosmoabc} was released \citet{akeret2015} also presented a Python package  for forward modelling  through PMC-ABC.} Python ABC package for astronomy\footnote{For similar tools in the context of biology and genetics, see e.g. \citet{liepe2010,oaks2014}.}. The package is structured so that the simulation, priors and distance functions are given as input to the main PMC-ABC sampler. In this context, users can easily connect  the ABC algorithm to their  own simulator and verify the effectiveness of the tool in their  own astronomical problems. The package also contains exploratory tools which help defining a meaningful distance function and consequently point to appropriate choices before the sampler itself is initiated.

We first guide the user through a very simple toy model, in order to clarify how the algorithm and the package work. Subsequently, as an example of cosmological application, we show how the machinery can be used to define credible intervals over cosmological parameters based on galaxy clusters catalogues. 
Simulations for this example were performed using the \textit{Numerical Cosmology} library \citep[{{\sc numcosmo}};][]{vitenti2014}.
The connection between {\sc cosmoabc} and {\sc numcosmo} is implemented as an independent module that can be easily adapted to other cosmological probes.

The outline of this article is as follows. In section \ref{sec:bayes}, we give an overview of Bayesian perspective and the ABC algorithm. \textsc{cosmoabc}  package is presented in section \ref{sec:cosmoabc} through a simple toy model. Section \ref{sec:abc_clust} describes in detail how to connect \textsc{cosmoabc} and \textsc{numcosmo} to obtain constrains over cosmological parameters from galaxy cluster number counts.
Our final remarks are presented in section \ref{sec:conclusion}.


\section{Bayesian approaches to parameter inference }
\label{sec:bayes}

Statistical inference on unknown parameters is often a primary goal of a physical experiment design. Although it is possible, and at times even desirable, to encounter some unpredictable behaviour among the outcomes of an experiment or a measurement, in many situations the  experimentation aims at  establishing  constraints over the parameters of a model. In other words, the desire is to use real-world data to check prevailing theories.   

In the Bayesian framework, the data are seen as the accessible truth regarding a given physical process and the model as a representation of our understanding of such process. This approach is data-centred and allows us to update the model whenever new information becomes available. 
In other words, our goal is to determine the probability of a model given the data, 
\begin{equation}
p(\pmb{\theta}|\mathcal{D}) = \frac{p(\mathcal{D}|\pmb{\theta})p(\pmb{\theta})}{p(\mathcal{D})},
\label{eq:bayes}
\end{equation}
where $\theta$ is the vector of model parameters, $\mathcal{D}$ the data set, $p(\pmb{\theta}|\mathcal{D})$ is called the \textit{posterior}, the prior, $p(\pmb{\theta})$, represents our initial expectations towards  the model and $p({\mathcal{D}})$ is a normalization constant.

In this context, the model parameters themselves are considered random variables and  each individual  measurement corresponds to one realization of them. Thus, once our prior is confronted with the data, the outcome is a posterior probability distribution function (PDF). Using the posterior distributions we can determine \textit{credible intervals}, which represent our uncertainty about the model parameters\footnote{Not to be confused with the frequentist definition \textit{confidence interval}, where the parameter values are considered fixed and therefore, there is no probabilistic interpretation associated to them.}.  For example, one may be interested in the most-probable region of values for certain parameters.


\subsection{Approximate Bayesian Computation}
\label{subsec:abc}

The ABC algorithm uses our ability to simulate the physical process under investigation to bypass the necessity of an unknown or computationally too expensive likelihood function. It is based on the following crucial elements:
\begin{itemize}
\item a simulator, or forward model,
\item prior probability distributions over the input parameters $p(\pmb{\theta})$,
\item a distance function, $\rho(\mathcal{D}_1,\mathcal{D}_2)$. 
\end{itemize}

As a simple example, consider the following toy model: a given physical process can be probed through a catalogue of $P$ observations, $\mathcal{D}=\{x_i,...,x_{P}\}$. Our model states that this process is driven by a random variable, $\mathcal{X}$, following a Gaussian distribution, $\mathcal{X} \sim \mathcal{N}(\mu_0, \sigma_0)$. Thus our goal is to identify credible intervals over $\mu_0$ and $\sigma_0$ based on $\mathcal{D}$. Moreover, our prior states  that $\mu_0 \in[\mu_{-},\mu_{+}]$ and $\sigma_0 \in [\sigma_{-},\sigma_{+}]$. Hereafter, we will denote the model parameters as $\pmb{\theta}=\{\mu,\sigma\}$.

The main idea behind the  ABC algorithm can be summarized in three main steps:

\begin{itemize}
\item draw a large number of parameter values, $\pmb{\theta}^{i}$, from the prior distribution, $p$,  

\item for each $\pmb{\theta}^i$ generate a simulation, $\mathcal{D}_{\rm S}^i$, and calculate the distance between the observed to the simulated catalogues, $\rho^i=\rho(\mathcal{D},\mathcal{D}_{\rm S}^i)$,

\item approximate the posterior probability distribution using the fraction of $\pmb{\theta}^i$'s with smallest associated distances.

\end{itemize}

The  above method has been modified and further developed in the last decade, generating some alternatives to the main algorithm (e.g. \citealt{sisson07,drovandi11,MarinEtAl2012,DelMoralEtAl2012,Ratmann2013}). One of them is presented below.

\subsection{Distance}

In the toy model described above, we can safely determine the distance, $\rho$,  between the measured catalogue $\mathcal{D}$ and a simulated one $\mathcal{D}_{\rm S}$ as 
\begin{equation}
\rho = {\rm abs}\left(\frac{\bar{\mathcal{D}}-\bar{\mathcal{D}_{\rm S}}}{\bar{\mathcal{D}}}\right) + {\rm abs}\left(\frac{\sigma_{\mathcal{D}}-\sigma_{\mathcal{D}_{\rm S}}}{\sigma_{\mathcal{D}}}\right),
\label{eq:dist_toy}
\end{equation}

\noindent where $\bar{\mathcal{D}}$ is the mean of all measurements in catalogue $\mathcal{D}$ and $\sigma_{\mathcal{D}}$ is its standard deviation. Equation \ref{eq:dist_toy} encloses important properties, which should be present in any ABC distance function: the distance between two identical catalogues is zero and the distance value increases steeply as parameter values get further from the fiducial ones. We emphasis that the choice of the distance function is a crucial step in the design of the ABC algorithm and the reader must check its properties carefully before any ABC implementation is attempted.

\subsection{Population Monte Carlo ABC}

\textsc{cosmoabc} uses the algorithm proposed by \citet{BeaumontEtAl2009}, where successive steps towards the posterior  are achieved by applying an importance (or weighted) sampling in the set of parameter values whose distances satisfy a given initial threshold.  

We begin by drawing $M$ values from the prior, called \textit{particles}, $\{\pmb{\theta}^i\}$ with $i \in [1,M]$, such that $M>>N$ ($N$ is the number of samples needed to characterize the prior). For each particle we generate a forward model (simulation) and  calculate the distance between synthetic and real catalogues $\rho^i=\rho(\mathcal{D},\mathcal{D}_{S}^i)$. From this large set, we keep only the $N$ particles with smallest $\rho^i$, which  constitute the first \textit{particle system}  ($\mathcal{S}_{t=0}$) and determine a distance threshold for the next iteration ($\epsilon_{t=1}$) as the $75\%$ quantile of all  $\rho \in \mathcal{S}_{t=0}$. In this initial step, we associate to each particle the same weight, $W_{t=0}^j=1.0/N$, for $j \in [1, N]$.

\begin{table}
\caption{Glossary for algorithm \ref{alg:PMC-ABC}.}
\label{tab:glossary}
\begin{tabular}{l c }
\hline
 Parameter  										& Description \\
\hline
$\mathcal{D}$ 										& Observed data set \\
$\mathcal{D}_{\rm S}$ 								& Simulated catalogue  \\
$M$ 												& Number of draws for the first iteration\\
$\mathcal{S}$ 										& Particle system\\
$N$ 												& Number of particles in $\mathcal{S}$\\
$t$ 												& Time-step (iteration) index \\
$K$ 												& Number of draws index\\
$W$ 												& Importance weights  \\
$\epsilon$ 											& Distance threshold  \\
$\Delta$ 											& Convergence criterion   \\
$\theta$ 											& Vector of model parameters  \\
$p(\cdot)$ 											& Prior distribution  \\
$\rho(\cdot,\cdot)$ 								& Distance function  \\
\multirow{2}{*}{$\mathcal{N}(\bullet;\theta, C)$} 	&  \multirow{2}{*}{\parbox{3.75cm}{\centering Gaussian PDF at $\bullet$ with  $\mu=\theta$, cov$=C$ }}\\
   &   \\
\hline
\end{tabular}
\end{table}

\begin{algorithm} 
\label{alg:PMC-ABC}
 \caption{PMC-ABC algorithm implemented in \textsc{cosmoabc}.} \label{alg:abc1}
 \KwData{$\mathcal{D} \longrightarrow$ observed catalogue.}
 \KwResult{ABC-posteriors distributions over the model parameters.}
 $t \longleftarrow 0$ \\
 $K \longleftarrow M$ \\
 \For { $J=1,\ldots, M$}
  {Draw $\theta$, from the prior, $p(\theta)$.\\
     Use  $\theta$ to generate ${\mathcal{D}_{\rm S}}$.\\
  Calculate distance, $\rho = \rho({\mathcal{D}_{\rm S}}, \mathcal{D})$.\\
  Store parameter and distance values, $\mathcal{S}_{\rm ini} \leftarrow \{\theta, \rho\}$}
  Sort elements in $\mathcal{S}_{\rm ini}$ by $|\rho|$.\\
  Keep only the $N$ parameter values with lower distance in $\mathcal{S}_{t=0}$.\\
  $C_{t=0} \longleftarrow$ covariance matrix from $\mathcal{S}_{t=0}$\\
  \For{$L=1,\ldots,N$}{
  $W_1^{L} \longleftarrow 1/N$} 
 \While{$N/K > \Delta$}{
 $K \longleftarrow 0$. \\
 $K_{*} \longleftarrow 0$. \\
 $t \longleftarrow t+1$. \\
 $\mathcal{S}_t \longleftarrow []$.\\ 
 $\epsilon_t \longleftarrow$  75$^{th}$-quantile of distances in $\mathcal{S}_{t-1}$.\\ 
 \While{{\rm len(}$\mathcal{S}_t${\rm )}$\quad < \quad$N}{
 $K_{*} \longleftarrow K_{*}+1$\\
 Draw $\theta_0$ from $\mathcal{S}_{t-1}$ with weights $\bar{W}_{t-1}$.\\
 Draw $\theta$, from $\mathcal{N}(\theta_0, C_{t-1})$.\\
 Use $\theta$ to generate $\mathcal{D}_{\rm S}$.\\
 Calculate distance, $\rho = \rho(\mathcal{D}_{\rm S}, \mathcal{D})$\\
 \If{$\rho\leq \epsilon_t$}{
 $\mathcal{S}_t \longleftarrow \{\theta,\rho, K_{*}\}$\\
 $K \longleftarrow K + K_{*}$\\
 $K_{*} \longleftarrow 0$}
 }  
 \For{$J=1,\ldots,N$}{
   $\tilde{W}_t^{J} \longleftarrow $ equation (\ref{eq:weight}).} 
 $W_t \longleftarrow$ normalized weights. \\  
 $C_t \longleftarrow $ weighted covariance matrix from $\{\mathcal{S}_t,W_t\}$.
 }
\end{algorithm}

In subsequent iterations, $t>0$, we perform an \textit{importance sampling} from $\mathcal{S}_{t-1}$: a popular 
technique where one can draw  from a proposal distribution 
and re-weight the particle system so it targets the desired posterior distribution.

The parameter vector resulting from this importance sampling, $\pmb{\theta}_{\rm try}$, is used to simulate a catalogue and calculate its distance to the observed data, $\rho_{\rm try}$. The parameter $\pmb{\theta}_{\rm try}$ is stored if $\rho_{\rm try} \leq \epsilon_{t}$. This process is repeated until a new set of $N$ parameter values satisfying the distance threshold is completed. For the new particle system, the weights are calculated as 
\begin{equation}
W_t^j=\frac{p(\pmb{\theta}_t^j)}{\sum_{i=1}^N W_{t-1}^{i}\mathcal{N}(\pmb{\theta}_t^j;\pmb{\theta}_{t-1}^i, C_{t-1})},
\label{eq:weight}
\end{equation}

\noindent  where $W_t^j$ denotes the weight associated to the $j-th$ particle in particle system $t$, $p(\pmb{\theta}_t^j)$ corresponds to the prior probability distribution calculated at $\pmb{\theta}_{t}^j$ , $W_{t-1}^i$ is the weight of the $i-th$ particle in particle system $t-1$ and $\mathcal{N}(\pmb{\theta}^j;\pmb{\theta}_{t-1}^j, C_{t-1})$ represents a Gaussian PDF\footnote{In general, the Gaussian PDF works well, but can be replaced with a different distribution if the parameter space has special restrictions, e.g. only takes integer values.} centred in $\pmb{\theta}_{t-1}^i$, with covariance matrix built from $\mathcal{S}_{t-1}$ and calculated at $\pmb{\theta}^j$. 

Once the new weights are determined, we start the construction of a new particle system and the algorithm is repeated until convergence. 
As pointed out by \citet{beaumont2009}, this is achieved when the ABC posterior no longer changes substantially with subsequent iterations. Here we consider that the system converged when the number of draws necessary to construct a  particle system is much larger than $N$ (see algorithm \ref{alg:PMC-ABC} and Section \ref{subsec:runABC}).  Each iteration brings us closer to the ``true'' PDF bypassing the need of a full likelihood calculation. Moreover, as the calculation of one particle is independent from the others within each iteration, the algorithm itself is more easily parallelizable than a standard MCMC. 


\section{\textsc{cosmoabc}}
\label{sec:cosmoabc}

In \textsc{cosmoabc}, our toy model can be represented by a simulation function,
\begin{lstlisting}
from scipy.stats import norm
import numpy as np

def my_simulation(v):
    """ Toy model simulator """ 
    
    dist = norm(loc=v['mean'], 
                scale=v['std'])
    l1 = dist.rvs(size=v['n'])
    
    return np.atleast_2d(l1).T 
\end{lstlisting}

\noindent where \texttt{v} is a dictionary of input parameters whose keywords \texttt{mean} and \texttt{std} determine the mean and standard deviation of the underlying Gaussian distribution, respectively,  and \texttt{n} denotes the total number of objects in the catalogue.   Analogously, a flat prior would be written as\footnote{The \texttt{func} argument is needed so we can retrieve a realization and the probability distribution itself. This is used by \textsc{cosmoabc} in the calculation of the weights.}
\begin{lstlisting}
from scipy.stats import uniform

def my_prior(par, func=False):
    """Flat prior"""
    
    gap = par['pmax'] - par['pmin']
    pdf = uniform(loc=par['pmin'], 
                  scale=gap)
    if func == False:
        draw = pdf.rvs()
        return draw
    else:
        return pdf
\end{lstlisting}

\noindent with \texttt{par} as a dictionary of input parameters and the keys \texttt{pmin} and \texttt{pmax} determining the boundaries of the distribution.

The distance function should be written as
\begin{lstlisting}
import numpy as np

def my_distance(d2, p):
    """Distance function."""
    
    dmean = np.mean(p['dataset1']) - 
            np.mean(d2))
    dstd = np.std(p['dataset1']) - 
           np.std(d2)        
           
    gmean = abs(dmean/
                np.mean(p['dataset1']))
    gstd = abs(dstd/
               np.std(p['dataset1']))
    
    rho = gmean + gstd
    
    return np.atleast_1d(rho)
\end{lstlisting}

\noindent and receive as input the simulated catalogue  \texttt{d2} and  the dictionary \texttt{p}. Notice that the observed catalogue is contained in \texttt{p}. So the distance to be calculated is between \texttt{p[\textquotesingle dataset1\textquotesingle]} and  \texttt{d2}\footnote{This format was chosen in order to optimize parallelization.}.

We must store these three functions in one file, $\texttt{<func\_file>}$, and edit the sample input file provided within \textsc{cosmoabc}. Each keyword in the sample input file is self-explanatory, so here we only emphasis the model and prior function parameters
\begin{lstlisting}
param_to_fit = mean std
param_to_sim = mean std n

mean_prior_par_name = pmin pmax
mean_prior_par_val = -2.0 4.0

std_prior_par_name = pmin pmax
std_prior_par_val = 0.1 5.0

mean_lim = -2.0 4.0
std_lim  = 0.1 5.0

mean = 2.0
std  = 1.0
n    = 1000
...
prior_func = my_prior my_prior
\end{lstlisting}

Notice that although the variables \texttt{mean} and \texttt{std} are free parameters, we need to provide an initial numerical value, within the constraints allowed by the prior.  The parameter \texttt{prior\_func} stores the prior PDF for all the free parameters, in the sequence declared in the variable \texttt{param\_to\_fit}. Such priors do not need to follow the same family of distribution. It is possible to define a flat prior for the first parameter and a Gaussian one for the second. In that case the user input file would include, for example, 
\begin{lstlisting}
mean_prior_par_name = pmin  pmax
...
std_prior_par_name = pmean pstd

...

prior_func = my_prior gaussian_prior
\end{lstlisting}
considering \texttt{pmean} and \texttt{pstd} as the mean and standard deviation for the Gaussian prior the second parameter under investigation.

\subsection{Visualizing distance behaviour}
\label{subsec:dist}

Before we attempt to use the ABC sampler, it is important to have an idea of how our distance definition behaves for different combination of model parameter values. \textsc{cosmoabc} has a tool which allows us to visually inspect the performance of our distance definition. The code randomly selects parameter values from the prior, performs the simulation and calculates the distance for each one of them. These distances are then plotted as a function of the parameter values, one parameter at a time. Ideally, the scatter of points in the $\rho \times \theta$ space should present a clear minimum in the neighbourhood of the most likely parameter value. 

In order to test a personalized distance function, do
\begin{lstlisting}
$ test_ABC_distance.py -i <input_file> 
                       -f <func_file> 
\end{lstlisting}

\begin{figure}[!t]
 \includegraphics[width=1\columnwidth]{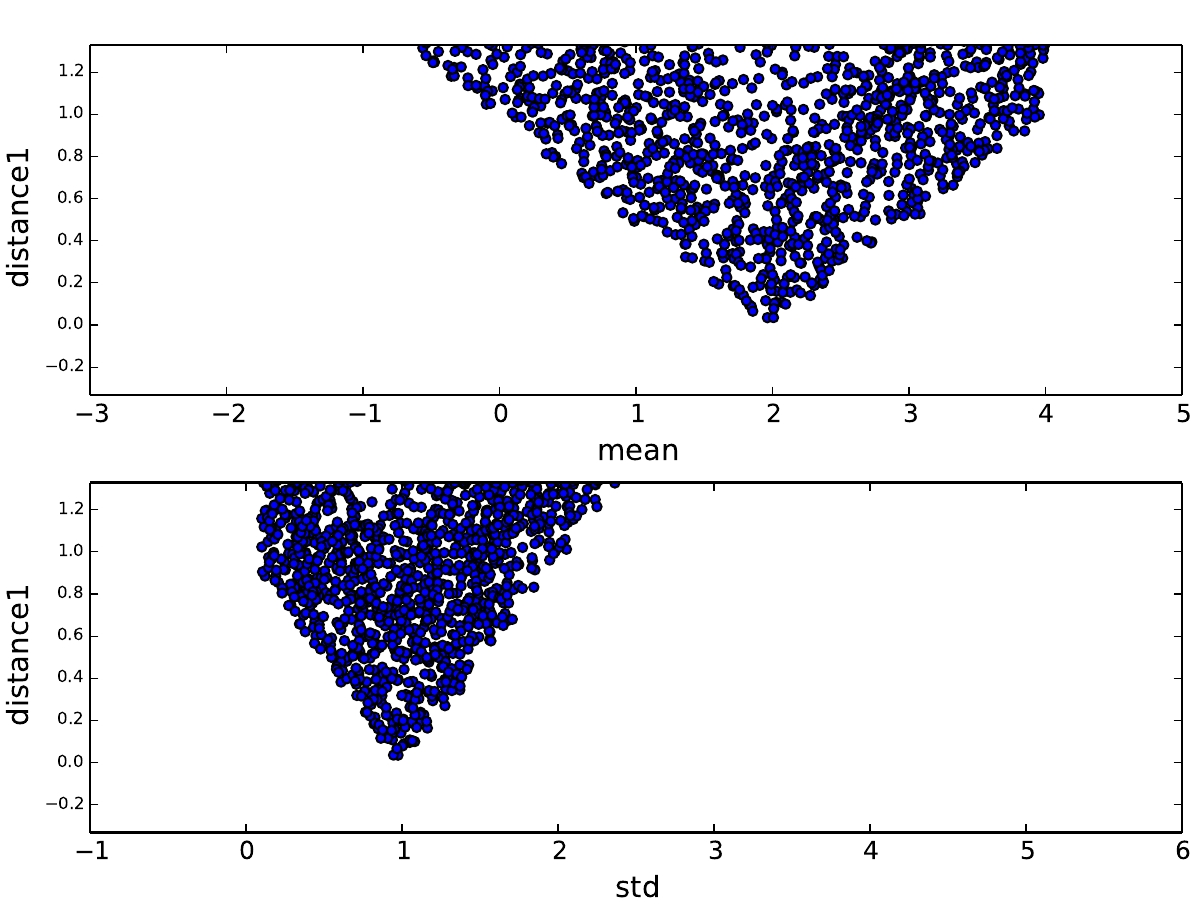}
 \caption{Behaviour of the distance function proposed in our toy model as a function of the free parameters \texttt{mean} (top) and \texttt{std} (bottom).}
\label{fig:distance_test}
 \end{figure}

\noindent An example of the result of this test  for the toy model we have been considering is shown in Figure \ref{fig:distance_test}. Notice that the distance behaves  as expected, approaching zero around the fiducial values \texttt{mean=2.0} and \texttt{std=1.0} and rapidly increasing as  parameter values move further away. 

It is worth mentioning that this procedure was implemented only to provide the user with an intuition regarding the distance function dependence with model parameters. The behaviour illustrated in Figure \ref{fig:distance_test} is a necessary but not sufficient characteristic of an ideal diagnostic. Selecting  an  appropriate distance function is an open and problem dependent challenge but it is an active area of statistical research \citep[see e.g.][]{fearnhead2012, blum2013, Ratmann2013}. A deeper investigation on the steps leading to an optimal distance definition, although very important, is out of the scope of this work.

\subsection{Running the ABC sampler}
\label{subsec:runABC}

After we are convinced of the performance of our distance function, we can proceed to the ABC sampler run. In \textsc{cosmoabc}, this is done through
\begin{lstlisting}
$ run_ABC.py -i <input_file> 
             -f <func_file>
\end{lstlisting}

The time necessary for the algorithm to converge depends on the efficiency of the simulator, the behaviour of the distance function and the number of particles in each particle system. We suggest an initial run with a fairly large convergence threshold, for example \texttt{delta = 0.25}. This means that the code will run until it is necessary to  take 4 times more draws than the number of particles in each particle system. In order to facilitate debugging and interaction with other codes, for each particle system \textsc{cosmoabc} outputs  parameter values, distance, distance threshold, computational time and weights for each particle in ASCII tables.

Once the algorithm converges, it is possible to visualize the results with
\begin{lstlisting}
$ plot_ABC.py -i <input_file>
              -f <func_file>
              -p T    
\end{lstlisting}
\noindent This will generate a file containing one snapshot for each particle system from \texttt{t=0} to \texttt{t=T}, as well as plots for the evolution of distance threshold, convergence criteria and computational time.  
From this first quick test, the user can either be satisfied with the achieved result or decide to continue iterating the sampler. If more iterations are required, it is only necessary to decrease the parameter $\texttt{delta}$ in the user input file and continue from the last completed particle system
\begin{lstlisting}
$continue_ABC.py -i <input_file>
                 -f <func_file>
                 -p T
\end{lstlisting}


\section{Case study: cosmological parameter inference from Sunyaev-Zeldovich surveys}
\label{sec:abc_clust}

The current concordance cosmology has been remarkably
successful in explaining the observed properties of large-scale
structures \citep{Tegmark2006,Benson2010}. In this framework, the formation of such structures proceeds in a hierarchical manner driven by pressureless
cold dark matter, where galaxy clusters stand out among the largest bound objects 
observed so far.   
The development of an underlying theory of 
cluster formation \citep[see][for a review]{Kravtsov2012}, allows us to use the abundance of clusters as well as their spatial distribution  as powerful cosmological probes \citep[e.g.,][]{Allen2008}.

There are, however, a couple of caveats which make this an interesting problem for the ABC approach: the model is not deterministic, in the sense that it considers the observed data as a realization of a Poisson distribution (analogously to the toy model studied before) and the unavoidable modelling of the observable uncertainties and errors in both, photometric redshifts and mass estimates (for a mathematical description we refer the reader to \ref{app:Cosmclust}, \citet{Penna2014} and references therein).  Using PMC-ABC surpasses the need to integrate a very complex likelihood function and reduces the influence of initial hypothesis on photometric redshift errors in the estimated posterior PDFs.

Since there is no  previous literature on the application of PMC-ABC to this particular problem, it is crucial to establish a proof of concept. Thus, here we present results from a completely synthetic framework, where the ``observed'' data, $\mathcal{D}$, is one instance of our forward model. This allows us to provide a controlled scenario and to ensure our capability of recovering the input parameter values. It also facilitates the identification (and quantification) of eventual biases in the final ABC-posteriors. 


\subsection{Simulations or the forward model}
\label{subsec:simul_in}

\begin{figure}[!t]
\includegraphics[width=\columnwidth]{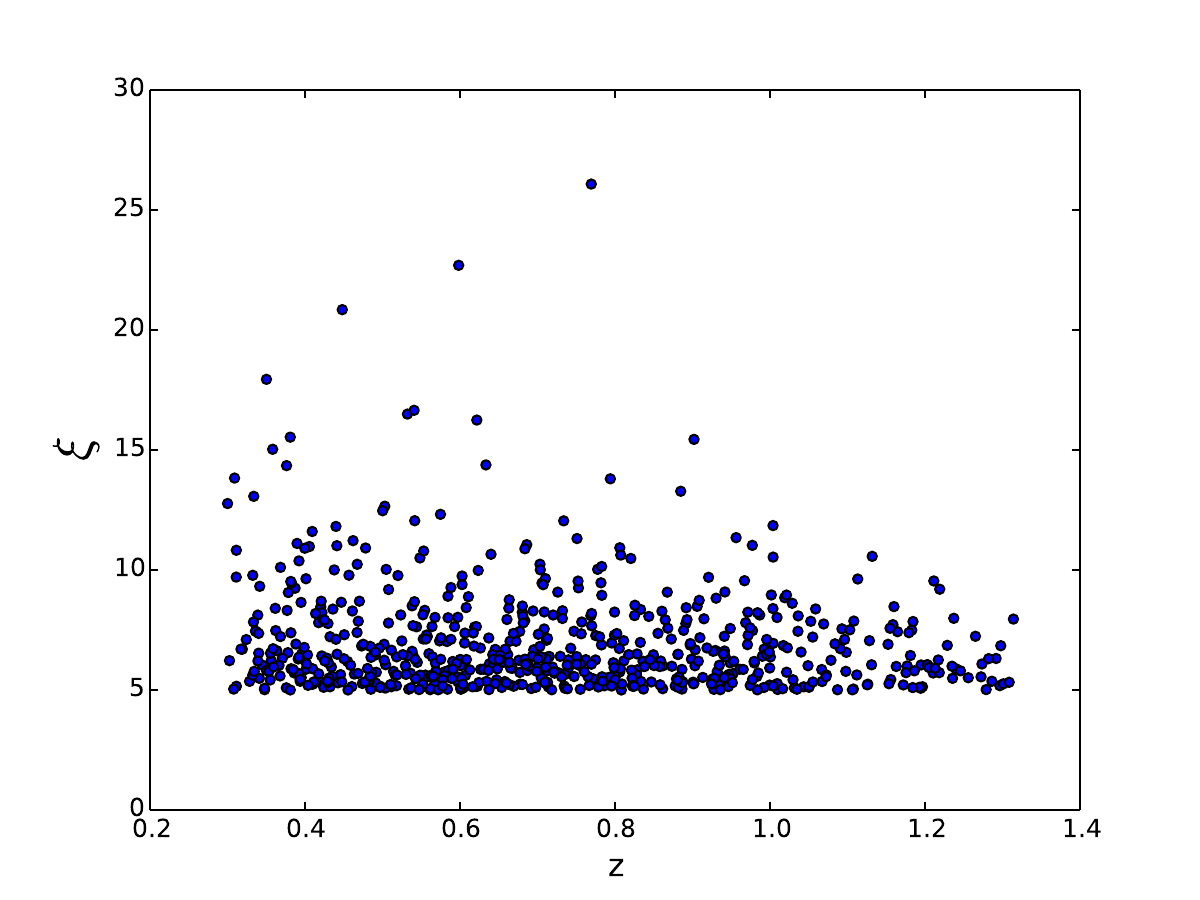}
\caption{Distribution of observed features (detection significance, $\xi$,  and redshift, $z$) in the ``observed'' catalogue, $\mathcal{D}$.
}
\label{fig:mockdata}
\end{figure}

Mock catalogues were generated with the {\sc numcosmo} library\footnote{\url{http://www.nongnu.org/numcosmo/}}, which provides a set of tools to perform cosmological calculations. The software allows a large range of possibilities for  input cosmological and astrophysical parameters as well as main survey specifications \citep[see][for a more detailed description]{vitenti2014}. Moreover, it can also account for the presence of uncertainties from photometric redshifts and mass-observable relation  (hereafter, $\xi$-mass relation, where $\xi$ is the detection significance) which are crucial for a coherent analysis of galaxy clusters number counts. 

Cosmological and astrophysical parameters for the fiducial model were chosen in accordance to \citet{reichardt2013}: $\Omega_c = 0.218$, $\sigma_8 = 0.807$, $w = -1.01$,  $\Omega_b = 0.044$, $H_0 = 71.15$ km/s/Mpc, $n_s = 0.97$,  $A_{SZ}=6.24$, $B_{SZ}=1.33$, $C_{SZ}=0.83$ and $D_{SZ}=0.24$ (see \ref{sec:SZ_mass} for definitions). Telescope characteristics follow the SPT design, with minimum and maximum redshifts given by $z_{\rm min}=0.3$, $z_{\rm max}=1.32$, respectively, and survey area $\Delta \Omega=2500$ deg$^2$ \citep{Bleem2014}.

The simulator begins assuming that the total number of galaxy clusters  with $z \in [z_{\rm min}, z_{\rm max}]$ and $\xi \in [\xi_{\rm min}, \infty)$ follows a Poisson distribution. 
It  then generates a realization of this distribution, 
$N_{\rm sim}$,  and the corresponding catalogue $\{\xi_i, z_i\}$, for $i \in\{1,N_{\rm sim}\}$ \citep[for details in the process see][appendix B]{Penna2014}.
Here, we investigate the three-dimensional space $\{\Omega_c, \sigma_8, w\}$ with flat initial priors, $\Omega_c \in [0.01,0.6]$, $\sigma_8 \in [0.5,1.0]$ and $w \in [-3.0,0.0]$. All other cosmological parameters are considered known and fixed at the values reported above.

\textsc{cosmoabc} contains a warp of the \textsc{numcosmo} simulator which can be accessed through the user input file keyword
\begin{lstlisting}
simulation_func = numcosmo_sim_cluster
\end{lstlisting}
and an example of the input file with all other options tailored for \textsc{numcosmo} simulations is also provided within the package.

Fig. \ref{fig:mockdata} displays the static simulated catalogue we used as ``observed''  data in the $\xi\times z$ sample space. The sample is composed by 671 clusters with   $z \in [0.30, 1.32]$  and $\xi \in [5, 26]$.  


\subsection{Distance}
\label{sec:summary}

\begin{figure*}[!t]
 \includegraphics[width=1\textwidth]{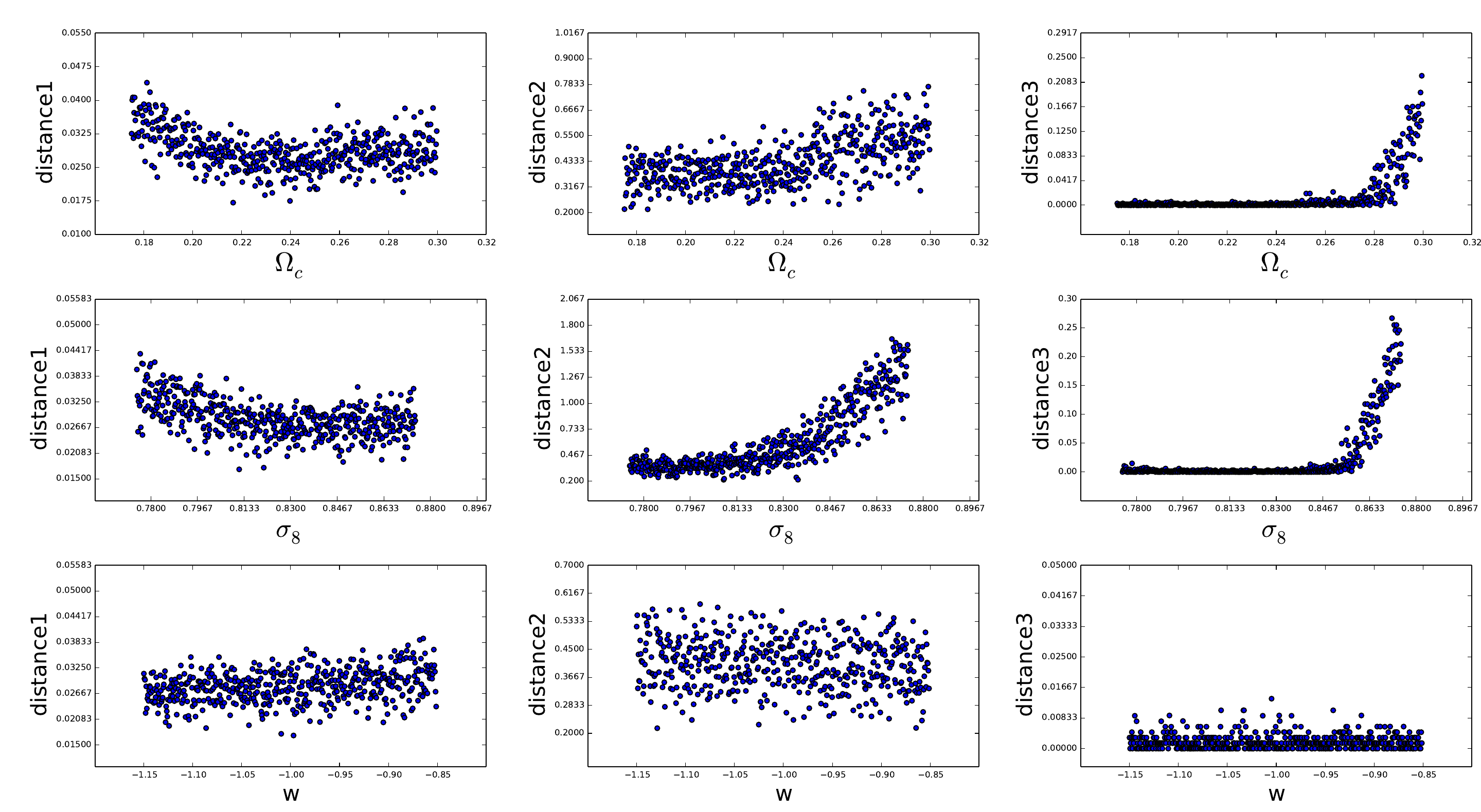}
 \caption{Behaviour of the quantile distance function in the context of galaxy clusters number counts. Each panel illustrates how the elements of the quantile based distance vary as a function of the cosmological parameters for $10^4$ random draws from the prior. Lines run through parameters and columns through distance elements.}
\label{fig:qdistance}
 \end{figure*}

The complexity enclosed in the cosmological simulations requires some sophistication in designing the distance function. \textsc{cosmoabc} has two built-in definitions which proved to be effective  in the galaxy cluster counts scenario: quantiles and Gaussian radial basis function (GRBF) distances.

The \texttt{distance\_quantile} function returns a vector $\pmb{\varrho}$,  having  $L+1$ dimensions,  where $L$ is the number of measured features\footnote{In our case, $L=2$, for observed features $\xi$ and redshift.}. For each  feature (column in the catalogue), it  calculates a few equally spaced quantiles\footnote{The number of quantiles if defined by the user in the input file, through the keyword \texttt{quantile\_nodes}.}. 
At every quantile, the values of the cumulative distribution functions (CDF) coming from simulated and observed catalogues are subtracted and the square root of their sum is returned. The last dimension accounts for the variability in the total number of objects. If $l$ is the number of objects in $\mathcal{D}$ and $l_S$ is the number of objects in $\mathcal{D_{S}}$, the last element of $\pmb{\rho}$ will be
\begin{equation}
\rho_{-1} = \max\left[{\rm abs}\left(1-\frac{l}{l_S}\right), {\rm abs}\left(1-\frac{l_S}{l}\right)\right].
\end{equation}

In the construction of the first particle system, the magnitude of this vector, $|\pmb{\rho}|$, is used to select the set of $N$ particles with smaller distances. Once the first particle system is constructed, the distance threshold $\pmb{\epsilon}$ will also be a $L+1$-dimensional vector. A new set of parameter values $\pmb{\theta}$ will only be accepted to populate the next particle system if it satisfies the 3 distance thresholds independently. 

We emphasis that the this  is only a simple  and computationally fast distance definition which proved to be efficient in this synthetic scenario of cosmological inference from galaxy clusters number counts for the illustrative purposes of this work. 
Whenever using ABC in a real data situation, the user must design a distance function which preserves these features for the  problem at hand (e.g., see Section 3.3 of \citet{cameron2012}).

Figure \ref{fig:qdistance} illustrates the effectiveness of this distance definition in determining the cosmological parameters based on SZ flux measurements. The distance calculations  were performed using the \textsc{cosmoabc} tool described in section \ref{subsec:dist}, however, in order to make the visualization lighter, we display binned results in all three free parameters. In each panel the horizontal axis was divided in 500 bins and each dot represents the smallest distance found in that bin for $10^4$ draws.  From Figure \ref{fig:qdistance} we see that the first (comparison of CDF over redshift) and second  (comparison of CDF over $\xi$)  distance elements  do present a local minimum around the fiducial values for $\Omega_m$ and $\sigma_8$, although the behaviour is much lighter than in the previously discussed toy model (Figure \ref{fig:distance_test}). The role of the third element (comparison between the total number of objects) is to impose an upper limit on the free parameter values, since this element increases steadily for $\Omega_m\geq 0.28$ and $\sigma_8\geq 0.86$. We also see that there is little hope in using this distance to constraint $w$, since there is no significant  change in behaviour for the three distance elements. 


\subsection{Results}
\label{sec:res}

\begin{figure*}[!t]
\begin{minipage}{\textwidth}
\begin{center}
\animategraphics[width=0.9\textwidth, controls,final]{0.75}{results-}{0}{9}
\end{center}
\end{minipage}
\caption{Results from coupling \textsc{cosmoabc} to the \textsc{numcosmo} simulator. Use the control bottoms to display the evolution of particle systems. Frames run from successive iterations of the PMC-ABC algorithm. \textbf{Upper panel}: two-dimensional representation of the ABC posteriors in each iteration. \textbf{Lower left panel}: evolution of the dark matter density profile. \textbf{Lower centre panel}: evolution of the posterior over $\sigma_8$. \textbf{Lower right panel}: evolution of the PDF profile over the dark energy equation of state parameter.}
\label{fig:numcosmo}
 \end{figure*}

\begin{figure*}
\begin{minipage}{0.45\textwidth}
\includegraphics[width=1\columnwidth]{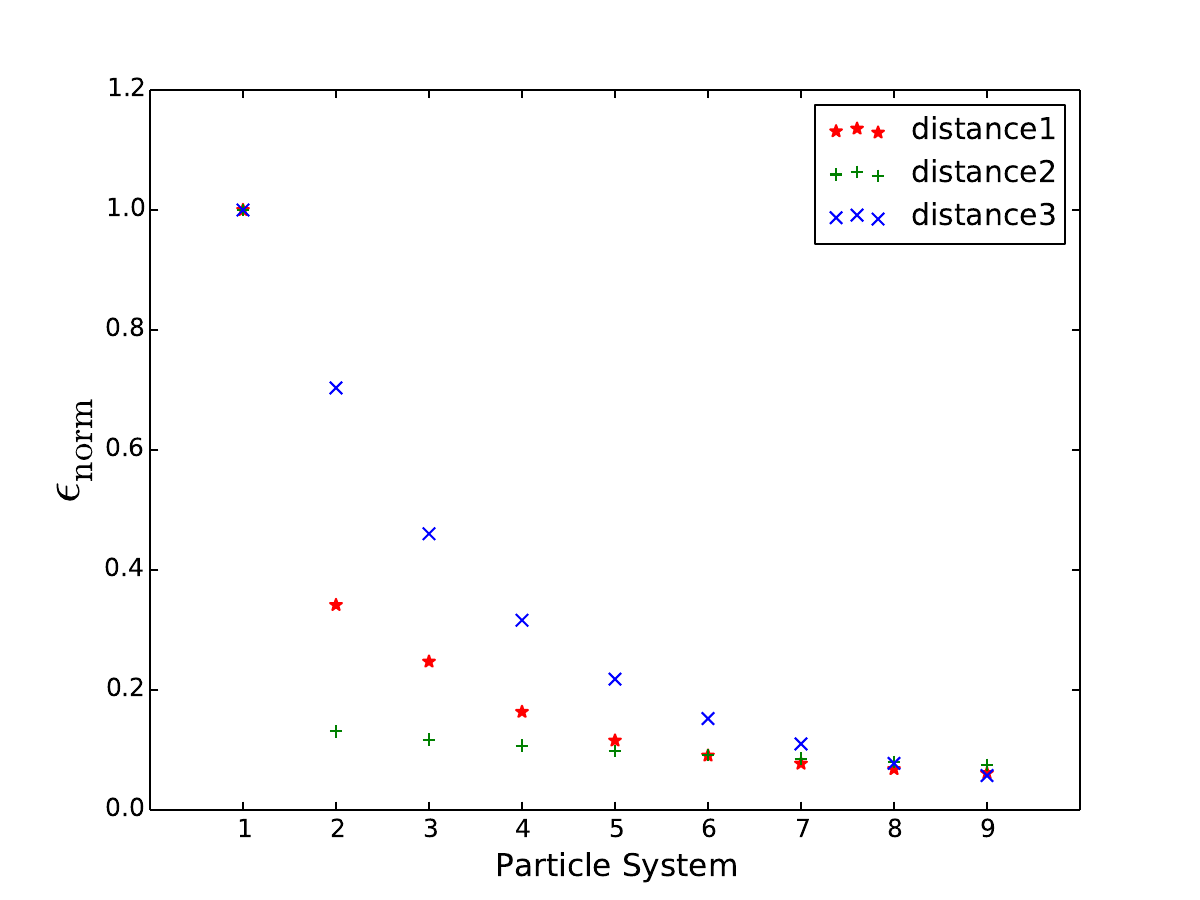}
\caption{Evolution of the distance threshold. The first (stars), second (+) and third (x) elements  of the quantile distance function were normalized by their respective larger values. The horizontal axis runs through 
all the particle systems shown in Figure \ref{fig:numcosmo}.}
\label{fig:epsilon_evol}
\end{minipage}
\qquad
\begin{minipage}{0.45\textwidth}
\includegraphics[width=\textwidth]{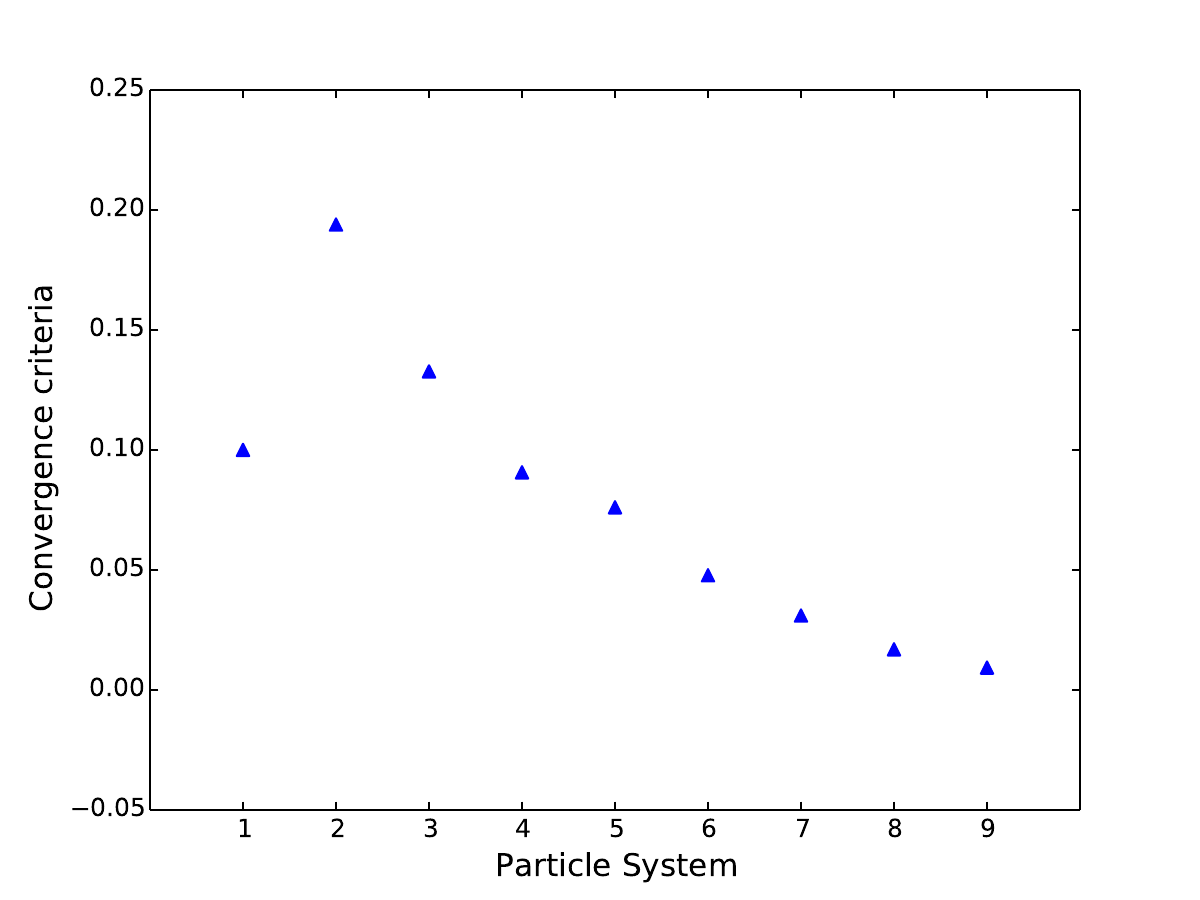}
\caption{Evolution of the convergence criteria for results shown in figure \ref{fig:numcosmo}.\vspace{0.75cm}}
\label{fig:delta_evol}
\end{minipage}
\end{figure*}

Specific tools  are also available for the case of a SZ survey using \textsc{numcosmo}. 
Once all the choices are made in the user input file it is possible to run the ABC sampler using
\begin{lstlisting}
$ run_ABC_NumCosmo.py -i <input_file>
\end{lstlisting}
 Analogously, if the user is interested in continuing the calculations from a given particle system \texttt{T} on, this can be done using
\begin{lstlisting}
$ continue_ABC_NumCosmo.py 
               -i <input_file>
               -p T
\end{lstlisting}
In case a user defined distance or prior is chosen, it is necessary to include the \texttt{-f} option followed by the name of the function file in both examples above.
Plots can be generated as shown in section \ref{subsec:runABC}.

Credible intervals  from the ABC-PMC estimated posterior distributions are shown in Figure \ref{fig:numcosmo}. The upper panels show 2-dimensions posteriors over the free parameters and the bottom panels display the profile of each parameter individually. Frames show the evolution of the approximated posteriors for consecutive particle systems. 
The first frame merely represents the initial prior: a flat PDF  over all the free parameters. The next frame displays results from the first particle system ($t=0$), where we generate a large number of simulations ($M=50000$) and kept only the 10\% with smaller distance. From $t=1$ on we clearly see how the posterior evolves and adapts through subsequent iterations.
The credible intervals not only shrink, but also become asymptotically well behaved for further particle systems.

Worth noting, for this example  each particle system holds $N=5000$ particles and the convergence was achieved in 9 iterations  for \texttt{delta = 0.01}.  In case a tighter posterior is desirable for $\Omega_c$ and $\sigma_8$, one can simply decrease the convergence criteria, letting the system evolve for a little longer. If further information on $w$ is desired, a more informative  distance definition should also be used\footnote{The quantile distance was chosen due to its simplicity and low computational cost. \texttt{cosmoabc} also contains a distance definition (\ref{ap:GRBF}) which accounts for potential  correlations between two parameters in a catalogue. We advise the user to consider the GRBF distance as well as the combination with other cosmological probes in case tight intervals over $w$ are desired.} (see Figure \ref{fig:qdistance}).
The evolution of the distance threshold and convergence criteria are shown in Figures \ref{fig:epsilon_evol} and \ref{fig:delta_evol}, respectively.


\section{Final remarks}
\label{sec:conclusion}

We presented \textsc{cosmoabc}, a Python implementation of Population Monte Carlo Approximate Bayesian Computation (PMC-ABC) algorithm with adaptive importance sampling.
 Traditional methods of parameter inference are useful if the likelihood is available and feasible to compute. Due to the increasing amount of data and their complex  modelling in all areas of astronomy and cosmology, more and more computational power is required in order to explore larger parameter spaces whose internal correlations can often be impractically complicated or unknown.  Thus, obtaining a statistical tool which bypasses the need of fully evaluating the likelihood is imperative.
PMC-ABC  presents an interesting alternative. \textsc{cosmoabc} is the first such implementation targeted to the astronomy and cosmology community.

The cosmological simulations are done through a connection with the \textit{Numerical Cosmology}  library (\textsc{numcosmo}), but the code is flexible enough for user-specified distance, simulation and prior functions. 

We stress that ABC is not a substitute for standard MCMC algorithms when the likelihood is completely known or easy to calculate. It is a viable alternative when we are not able to handle the likelihood itself, and thus in situations where a MCMC is not feasible.  

In this work, we demonstrated the power of \textsc{cosmoabc} in estimating posterior probability distributions in two situations: a simple toy model and a complex cosmological simulation of a Sunyaev-Zeldovich survey. In both cases, we demonstrated how \textsc{cosmoabc} allows a good approximation of the true posterior probability distribution  with a fairly simple and user-friendly interface. We used a completely synthetic environment in order to demonstrate the efficiency of the method and to be able to address the accuracy of the results.  We hope this will be useful not only to cosmologists, but to all research areas in astronomy where simulations are becoming increasingly more accessible and systematics are making  likelihood functions even more intractable.

The code is published under GPLv3 in PyPl and Github and documentation can be found in Readthedocs\footnote{\url{http://cosmoabc.readthedocs.org/en/latest/}}.


\section*{Acknowledgements}

We are happy to thank Alberto Krone-Martins, Christian Robert, Jonnathan Elliot, Joseph Hilbe and Madhura Killedar for insightful discussions and suggestions. 
This work is a product of the first COIN Summer Residence Program (August/2014, Lisbon).
EEOI, RSS, AMMT and VCB thank the SIM Laboratory of the \emph{Universidade de Lisboa} for hospitality during the development of this work. The IAA Cosmostatistics Initiative (COIN)\footnote{\url{https://asaip.psu.edu/organizations/iaa/iaa-working-group-of-cosmostatistics}} is a non-profit organization whose aim is to nourish the synergy between astrophysics, cosmology, statistics and machine learning communities.
This work was partially supported by the ESA VA4D project (AO 1-6740/11/F/MOS). EEOI is partially supported by the Brazilian agency CAPES (grant number 9229-13-2). SDPV is supported by CAPES (grant number 2649-13-6).
MPL acknowledges the Brazilian agency CNPq (PCI/MCTI/INPE program and grant number 202131/ 2014-9) for financial support. 
VCB was supported by CNPq-Brazil, with a fellowship within the program Science without Borders, FAPESP (grant number 2014/21098-1) and CAPES. AMMT was supported by the FCT/IDPASC grant number SFRH/ BD/ 51647/2011.

Work on this paper has substantially benefited from using the collaborative website AWOB\footnote{\url{http://awob.mpg.de}} developed and maintained by the Max-Planck Institute for Astrophysics and the Max-Planck Digital Library. This work was written on the collaborative \texttt{Overleaf} platform\footnote{\url{www.overleaf.com}}, and made use of the GitHub\footnote{\url{www.github.com}} repository web-based hosting service and \texttt{git} version control software.


\appendix


\section{Cosmology with galaxy cluster number counts}
\label{app:Cosmclust}

The model assumes that the number density of halos with mass in the range $[M, M + dM]$ is strongly linked to both the matter-energy content of the Universe and the statistical properties of the initial linear density contrast field \citep{Press1974, Bond1991, Sheth1999}. Hence, the comoving number density of dark matter halos can be written as 
\begin{equation}
 \frac{dn(M, z)}{d\ln M} = -\frac{\rho_m(z)}{M} f(\sigma_{R}, z) \frac{1}{\sigma_{R}} \frac{d\sigma_{R}}{d\ln M},
 \label{eq:mass_function}
\end{equation}
where $\rho_m(z)$ is the mean matter density at redshift $z$, $f(\sigma_{R}, z)$ is the multiplicity function \citep{Tinker2008}, and $\sigma^2_{R}$ is the variance of the linear density contrast filtered on the length scale $R$, 
\begin{equation}
\sigma^2_R(z) = \int_0^\infty \frac{dk}{2\pi^2} k^2 \, P(k, z) \vert W(k,R) \vert^2,
\label{eq:var:R:z}
\end{equation}
with  a spherical top-hat window function, $W(k, R)$, and the linear power spectrum given by
\begin{equation}\label{eq:powspec}
P(k, z) = A \, k^{n_s} \, T(k)^2 \, D(z)^2.
\end{equation}
In the expression above the power spectrum depends on the  spectral index, $n_s$, the linear growth function, $D(z)$, normalized such that $D(0) = 1$, and on the transfer function, $T(k)$  \citep{Eisenstein1998}. The normalization factor  is written as
\begin{equation}\label{eq:norma_sigma8}
A = \frac{\sigma_8^2}{\int_0^\infty \frac{dk}{2\pi^2} k^{(n_s + 2)} T(k)^2 W^2(k,8)}.
\end{equation}

In order to obtain the mean number of DM  halos with mass in the range $[M, M + dM]$ and in the redshift interval $[z, z + dz]$, 
we combine the mass function 
with the comoving volume element $dV/dz$, namely,
\begin{equation}\label{eq:dN_dz}
\frac{d^2N}{dz d\ln M} dzd\ln M =  \frac{dV}{dz} \frac{dn(M, z)}{d\ln M} dzd\ln M.
\end{equation}
Assuming a flat universe,
\begin{equation}
\frac{dV}{dz} = \Delta\Omega \left(\frac{\pi}{180}\right)^2 \frac{c}{H(z)} \left(\int_0^z dz^{\prime} \frac{c}{H(z^{\prime})}\right)^2,
\label{eq:vol}
\end{equation}
where $\Delta\Omega$ is the survey area in square degrees and $H(z)$ is the Hubble parameter 
\begin{equation}\label{eq:hubble}
H(z) = H_0 \big[\Omega_m(1 + z)^3 + (1 - \Omega_{\rm m}) \, (1 + z)^{3(1 + w)} \big]^{1/2},
\end{equation}
with $\Omega_{\rm m}$ representing the fraction of total energy density in the form of 
matter ($\Omega_m$), $H_0$ being the Hubble constant and $w$ the equation of state of dark energy, considered to be constant.

This framework allows us to relate the mean number of DM halos within a certain range of mass and redshift (equation \ref{eq:dN_dz}) to the parameters describing the underlying cosmological model (equations \ref{eq:mass_function} to \ref{eq:hubble}). However, we still need to connect  the theoretical   redshift $z$ and  mass $M$ with   their equivalent   observable  quantities.
We begin by taking into account the uncertainties from photometric redshift, $z_{phot}$,  determination.

We  assume that $z_{phot}$ follows a Gaussian distribution with mean equal to $z$ and standard deviation $\sigma = 0.05(1 + z)$, which we refer to as $P(z_{phot} | z)$. 
Thus, the expected number of clusters for a given interval of $z_{phot}$ and $M$ can be written as,

\begin{equation}
\label{eq:d2N:phot:xi}
\frac{d^2N(M, z_{phot}, \pmb{\theta})}{dz_{phot} d\ln M} = \int dz P(z_{phot} | z) \frac{d^2N(M, z, \pmb{\theta})}{dz d\ln M},
\end{equation}%
where  $\pmb{\theta}$ comprises both the cosmological and astrophysical   parameters (such as those of the mass- observable relation - see \ref{sec:SZ_mass}). 

Estimating the mass enclosed in a given galaxy cluster is not a trivial task \citep[see e.g., ][and references therein]{Lagana2010,Giodini2013}. Traditionally, one requires the recognition of indirect signatures carrying such information into observable quantities, such as optical and X-ray emissions \citep{birkinshaw1999,carlstrom2002}. 
In particular, we use a mass-observable relation derived by the SPT team, which relies on measurements of the SZ effect.


\subsection{Cluster mass estimate from Sunyaev--Zeldovich effect}
\label{sec:SZ_mass}

The intracluster medium (ICM) is a hot plasma which interacts with photons of the cosmic microwave background (CMB) via Compton scattering, causing a spectral distortion in the CMB radiation. This is known as the SZ effect.
The integrated thermal SZ flux, $Y_{SZ}$, is proportional to the total thermal energy of the ICM \citep{Barbosa1996, Motl2005} and consequently it is possible to use the SZ distortions on the CMB to estimate the mass of the cluster. 

Due to significant uncertainties in the direct determination of $Y_{SZ}$, we  follow here  the strategy reported by the SPT \citep{Vanderlinde2010, Benson2013, reichardt2013}, where an unbiased estimator $\zeta$ of the detection significance (signal to noise ratio) $\xi$  is used as a mass proxy.
In this context, $\zeta = \sqrt{\langle\xi\rangle^2 - 3}$ and $\zeta \propto Y_{SZ}/N_{int}$, with $N_{int}$ denoting the noise per resolution element or the integrated noise over several resolution elements for  unresolved and resolved detections, respectively. Moreover, the adopted  mass scaling relation is given by  

\begin{equation}\label{eq:zeta_to_mass}
\zeta = A_{SZ} \left( \frac{M_{500}}{3 \times 10^{14} M_\odot h^{-1}} \right)^{B_{SZ}} \left( \frac{E(z)}{E(0.6)} \right)^{C_{SZ}},
\end{equation}
where $E(z)=H(z)/H_0$, $M_{500} = (4\pi/3)\, 500\rho_{\rm crit} \, R^3_{500}$, with $\rho_{\rm crit} = 3H_0^2/8\pi G$   as the critical energy density,  $R_{500}$ the radius enclosing 500$\times \rho_{\rm crit}$ at the cluster redshift, and the scaling relation parameters $A_{SZ}$ ($\zeta$-mass normalization), $B_{SZ}$ ($\zeta$-mass slope) and  $C_{SZ}$ ($\zeta$-mass redshift evolution) can be determined concomitantly  with the cosmological parameters. 
Finally, substituting the true mass by the unbiased estimator in equation \ref{eq:d2N:phot:xi}, the number of clusters with $\xi \in [\xi, \xi + d\xi]$ and $z_{phot} \in [z_{phot}, z_{phot}+dz_{phot}]$ 
can be expressed as
\begin{eqnarray}
\label{eq:d2N:xi:phot}
\frac{d^2N(\xi, z_{phot}, \pmb{\theta})}{dz_{phot} d\xi}&&  =  \int dz P(z_{phot} | z) \nonumber \\
&&\int d\ln M\int d\zeta\frac{d^2N(M, z, \pmb{\theta})}{dz d\ln M} \nonumber \\  
&& P(\xi | \zeta) \, P(\ln\zeta | \ln M).\nonumber\\
&&
\end{eqnarray}
Following \citet{Benson2013}, and  \citet{reichardt2013}, we assume
\begin{eqnarray}
P(\ln\zeta | \ln M) d\ln \zeta&&=\frac{1}{\zeta\sqrt{2\pi}D_{SZ}} \times \nonumber \\ &&\exp\left[{-\frac{(\ln \zeta - \ln M)^2}{2 D_{SZ}^2}}\right] d\zeta, \nonumber \\ 
&&
\end{eqnarray}
with $D_{SZ}$ being the log-normal scatter in $\zeta$, and 
\begin{equation}
P(\xi | \zeta) d\xi = \frac{1}{\sqrt{2\pi}} \exp\left[{-\frac{\left(\xi - \sqrt{\zeta^2 + 3}\right)^2}{2}}\right] d\xi. 
\end{equation}


\section{Distance based on Gaussian Radial Basis Function}
\label{ap:GRBF}

We describe bellow a distance defined in terms of Gaussian radial basis functions (GRBF), $\rho_{\rm GRBF}$ (within \textsc{cosmoabc} the function is called \texttt{distance\_GRBF}). This is not a distance in the mathematical sense, since $\rho_{\rm GRBF}\left(\mathcal{D}, \mathcal{D}_S\right) \neq \rho_{\rm GRBF}\left(\mathcal{D}_S, \mathcal{D}\right)$, but as the PMC-ABC is centred in the observed catalogue, it is enough to guide the sampler to the correct posterior distribution over the evolution of particle systems.

For a given simulated sample, we compute the approximation of its underlying model using 
a GRBF  interpolation 
\citep{Tsybakov2008},
\begin{equation}
\label{eq:GRBF}
G_{\rm R}(\xi,z|\mathcal{D})=\sum_{i=1}^{N_{\rm sim}}\frac{1}{2\pi\sqrt{\det(\bm{C})}}\exp\left(-\frac{\bm{d}_i^T\cdot\bm{C}^{-1}\cdot\bm{d}_i}{2}\right),
\end{equation}
where 
\begin{eqnarray}
\bm{d}_i &=& (\xi - \xi_{\mathcal{D}_S}^i, z - z_{\mathcal{D}_S}^i),\nonumber\\ 
\bm{C} & = & s^2\rm{cov}(\mathcal{D}), \nonumber
\label{eq:cov}
\end{eqnarray}
$s$ is a scale parameter and $\rm{cov}(\mathcal{D})$ is the covariance matrix of the observed data.
Each element of the sum in equation (\ref{eq:GRBF}) works as a kernel density distribution centred in the $i-th$ simulated cluster with $(\xi_i$, $z_i)$. Consequently, $G_{\rm R}(\xi_*,z_*)$ is the number of clusters we expect to observe having $(\xi_*,z_*)$.

We can re-obtain the total number of objects in the catalogue through 
\begin{equation}
\label{eq:ns}
\int\mathrm{d}\xi\int\mathrm{d}z G_{\rm R}(\xi,z\vert \mathcal{D}) = N_{\rm sim}.
\end{equation}
Making use of the auxiliary function 
\begin{equation}
\label{eq:f_rho}
\ln f (\mathcal{D}\vert \mathcal{D}_S) = \sum_{j=1}^{N}\ln(G_{\rm R}(z_j,\xi_j\vert \mathcal{D}_S)) - N_{\rm sim},
\end{equation}
with the index $j$ running through all the data points in $\mathcal{D}$ \citep{Penna2014},  the distance between  the ``observed'' and simulated catalogues is given by 
\begin{equation}
\label{eq:dist}
\rho(\mathcal{D},\mathcal{D}_S) = -2\ln\left(\frac{f(\mathcal{D}\vert \mathcal{D}_S)}{f(\mathcal{D}\vert \mathcal{D})}\right).
\end{equation}
Therefore, in each iteration, $t$, we will only accept those parameter values whose forward  model satisfy $\rho_t(\mathcal{D},\mathcal{D}_S) \leq \epsilon_t$ and the ABC-PMC algorithm can be employed normally\footnote{Note that the denominator in equation  \ref{eq:dist} is  a scalar independent of the mock data. This  ensures the distance function convergence.}.

The scale parameter $s$ (equation \ref{eq:cov}) regulates our tolerance towards distinct distributions which produce the same total number of objects in a catalogue. 
Suppose  we begin with a large $\epsilon_1$ and follow the ABC-PMC algorithm reducing $\epsilon_t$ at each time-step. If $s$ is too small the probability of finding parameter values satisfying the distance threshold will drop steadily, rendering the algorithm unable to further reduce $\epsilon_t$. On the other hand, if $s$ is too large, the density function will evolve to a very flat behaviour losing all information about the underlying distribution of $\mathcal{D}$.  Thus, $s$ must be chosen such that most of the shape information in $\mathcal{D}$ is retained, while still being feasible to reduce $\epsilon_t$ until the desired precision is achieved.
For the specific case  outlined here, 
we found that any $s\in [0.1, 0.5]$  will lead to well constrained and unbiased results. 

In order to call this distance definition from within \textsc{cosmoabc}, the input file must include the extra parameter \texttt{s} and the function definition
\begin{lstlisting}
...
distance_func = distance_GRBF
s  = 0.15 
...
\end{lstlisting}

The GRBF distance is  more time consuming, since it takes into account the correlation between the observed features. However, it might be worth to use it in highly correlated data scenarios.
The current version of the \textsc{numcosmo} library includes an ABC sampler using  the GRBF distance.


\section*{REFERENCES}
\bibliographystyle{mn2e}
\bibliography{ref}

\end{document}